\documentclass[12pt]{book}

\usepackage[dvips]{graphicx,color}
\usepackage{makeidx,tsukuba}

\makeauthorindex
\makeindex

\begin{document}

\BookTitle{\itshape The 28th International Cosmic Ray Conference}
\CopyRight{\copyright 2003 by Universal Academy Press, Inc.}
\pagenumbering{arabic}

\chapter{UHECR Anisotropy from Luminous Infrared Galaxies - Predictions for the
Pierre Auger Observatory }

\author{Giller Maria,
Michalak Wojciech  and Smialkowski Andrzej}
{\it  Division of Experimental Physics, University of
Lodz, Pomorska 149/153, 90-236, Lodz, Poland, e-mail:
asmial@kfd2.fic.uni.lodz.pl}

\section*{Abstract}
We consider the hypothesis that luminous infrared galaxies (LIRGs) are sources
of the UHECRs. By associating the AGASA triplet with the Arp 299 galaxy we
obtain reasonable values for Galactic and extragalactic magnetic fields. We
predict what the southern sky, to be seen by the Auger experiment, should look
like, so that the LIRG hypothesis could be verified soon.
 \section{Introduction}
In the quest for the origin of the ultra high energy cosmic rays (UHECR) various
methods have been pursued, where the most promising one is to look for the
particle point sources. Some directional correlations with quasars have been
found [4,8] but we are rather sceptical about considering these very distant
objects as sources of UHECR particles, because of the GZK effect. This paper is
partly a continuation of our previous hypothesis that UHECRs may come from
luminous infrared galaxies (LIRGs) [6]. These objects have IR emissivities
larger by an order of magnitude than normal galaxies.  Being mainly colliding
galaxies they may constitute favourable conditions for high energy particle
acceleration (see also [1]). In [6] we have shown that there is quite a
reasonable directional correlation between $(4-8)\times10^{19}eV$ AGASA showers
and LIRGs on the northern sky. In particular, the AGASA triplet [7] coincides
well with the brightest (up to 70 Mpc) extragalactic IR source, Arp 299, 42 Mpc
away. Here, we analyse further the possibility of Arp 299 being the source of
the triplet particles, allowing for their changing directions in extragalactic
and Galactic magnetic field. We also predict what  the southern sky would look
like in the highest  energy cosmic particles, as measured by the Pierre Auger
Observatory in Argentina [3]. \section{AGASA triplet from Arp 299 ?}
\begin{figure}
\includegraphics[width=0.4\textwidth,height=6.05cm]{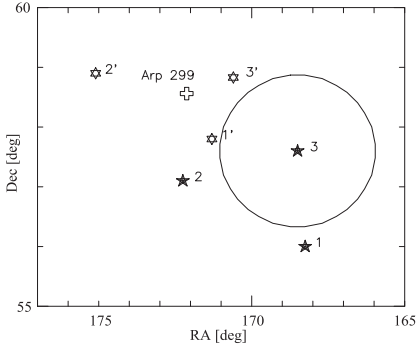}
    \hfill
    \begin{minipage}[b]{0.48\textwidth}
  \caption{Close look at the sky near Arp 299 (cross). Dark stars denote AGASA
triplet, open stars - particle directions before entering the Galaxy. Circle
shows uncertainty of shower direction $\sim 1.5^{\circ}$ above
$5\times10^{19}$eV. }\label{fig:1} \end{minipage}
\end{figure}

The positions of the three shower directions of the AGASA triplet, together with
the direction towards Arp 299 are shown in Fig \ref{fig:1}. The particle
energies are  54, 55 and 78 EeV correspondingly for shower 1,2 and 3.
As the distance to Arp is 'only' about 42 Mpc, let us assume that these
particles may have arrived from this source with negligible energy losses and
follow the consequences.\\ \indent 
As all three particles  are
shifted more or less in similar directions away from the source it may be that
it is the Galactic magnetic field responsible for that. One can find such a
value of the integral $\int B_{\perp}dl$ (along the line of sight) that the
combined occurence of the remaining particle deflections in the extragalactic
chaotic fields is  most probable. Let us denote the deviation of the
particle arrival direction from that towards the source by $\vec{r'}$. If it is
caused by multiple independent small deviations in the extragalactic magnetic
field, then the probability that three particles will have deviations
$\vec{r'_{1}}$,$\vec{r'_{2}}$ and $\vec{r'_{3}}$, each within a small solid
angle $d\Omega_{1}$,$d\Omega_{2}$ and $d\Omega_{3}$, equals:
\begin{equation}
dp(\vec{r'_{1}},\vec{r'_{2}},\vec{r'_{3}})=\prod_{i=1}^{3}\frac{1}{\sigma_{i}^
{2}}\,exp\Big(-\frac{\vec{r'^{2}_{i}}}{\sigma_{i}^{2}}\Big)d\Omega_{i}
\label{eq:1}
\end{equation}
where $\sigma_{i}$ is the r.m.s. of  $r'_{i}$ and equals:
$\sigma_{i}=\sqrt{2}\,q_{i}B_{rms}\sqrt{lD}/(3E_{i})$
(for a model with magnetic cells of size $l$), and $D$ is the distance to the
source.\\ \indent The Galactic field shifts the particle arrival
directions by: \begin{equation}
\vec{g_{i}}=\frac{q_{i}}{E_{i}}\int \vec{B}\times
\vec{dl}=\frac{0.53^{\circ}}{\epsilon_{i}}\int\vec{B_{6}}\times\vec{dl}\equiv
\frac{\vec{a}}{\epsilon_{i}}
\label{eq:2}
\end{equation}
where $\epsilon_{i}=E_{i}/10^{20} eV$, $B_{6}=B/10^{-6} G$ and $q_{i}$ is
the particle charge. We assume here that all particles have the same charge
$q=1e$. If the \emph{observed} deviations from the source
are $\vec{r_{i}}$ then $\vec{r'_{i}}=\vec{r_{i}}-\vec{g_{i}}$.
Introducing all this into (1) we obtain the
probability $dp$ as a function of $\vec{a}$ and $\sigma_{1,2,3}$. We want to
find such a vector $\vec{a}$ for which the probability $dp$ has a maximum value.
Differentiating (\ref{eq:1}) with respect to the two coordinates $a_{x}$ and
$a_{y}$ (where $x$ corresponds to declination and $y$ to
right ascension) and putting the derivatives equal to zero we obtain:
\begin{equation}
a_{x}=\frac{1}{3}\sum_{i=1}^{3}\epsilon_{i}\Delta x_{i}\quad and \quad
a_{y}=\frac{1}{3}\sum_{i=1}^{3}\epsilon_{i}\Delta y_{i}
\label{eq:6}
\end{equation}
where $\Delta x_{i}$ and $\Delta y_{i}$ are the components of  $\vec r_{i}$
Adopting the particular
values of $\Delta x_{i}$ and $\Delta y_{i}$ for the AGASA triplet with Arp 299
as the source we find that $\int B_{\perp}dl\simeq 2.4 \mu G\cdot kpc$. This
value does not look unreasonable for the Galactic latitude $\simeq 56^{\circ}$.
\\ \indent The particle new directions, after shifting by
$-\vec{a}/\epsilon_{i}$, are shown in Fig \ref{fig:1} by the open stars. The
remaining deviations $\vec{r'_{i}}$ depend on the parameter $b=B_{rms}\sqrt{l}$,
which can be found by maximizing again the combined probability (\ref{eq:1}).
So far we have neglected the experimental uncertainties of the shower
directions. For the AGASA triplet, however,  $\sigma_{i}\leq\sigma_{exp}$, so
that at this stage any conclusions about $B_{rms}\sqrt{l}$ have large
uncertainty. Nevertheless for $r'\leq 1^{\circ}$ one gets $b\leq
0.3\,nG\,Mpc^{1/2}$, a value consistent with the current
knowledge.\\ \indent
One can go even further and assume that all
particles in the multiplet were emitted at the same time i.e. the time interval
of their emission is much smaller than the time interval of their arrivals.
If the latter ($\simeq 3$ yrs) is compared to the r.m.s. of the arrival time
differences
\begin{equation}
\sigma_{\Delta t}\simeq
1.25\times10^{4}yrs\cdot(D/30Mpc)^{2}\cdot
(b/1nG\,Mpc^{1/2})^{2}\cdot\epsilon^{-2}
\end{equation}
then $b\simeq 10^{-2}\,nG\,Mpc^{1/2}$. It is much smaller than that estimated
from the above analysis of the deviation angles, if $\sigma_{exp}$ are
neglected. \section{UHECR sky predicted for the Auger
experiment} \begin{figure}[t]
    \includegraphics[width=14cm]{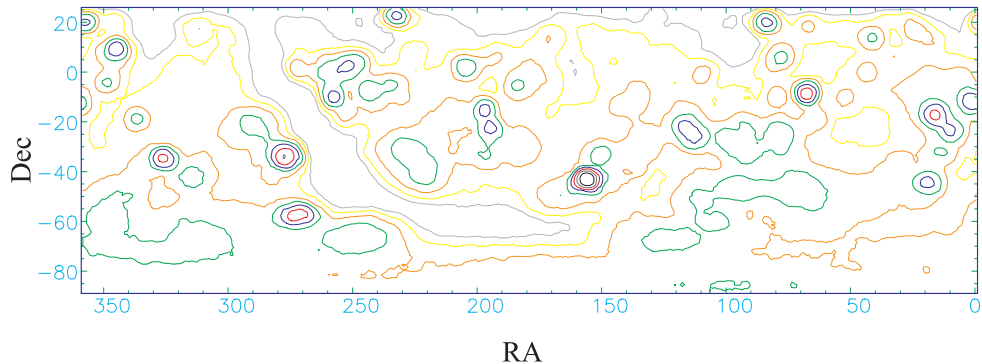}
  \caption{UHECR sky (energy above 40 EeV) predicted for Auger. Isolines (flux
per unit solid angle) spaced with factor 0.5.}
\label{fig:2}
\end{figure}
Assuming that LIRGs are the UHECR sources one can predict the angular
distribution of the shower directions as would be observed by the southern site of
the Pierre Auger Observatory. We assume that the cosmic ray
production rate of a LIRG is proportional to its IR emissivity. We adopt
the production energy spectrum $\propto E^{-2}dE$ up to $E_{max}=10^{21}eV$. In
the extragalactic space the particles are being scattered by the magnetic fields
such that $B_{rms}\cdot\sqrt{l}=1\,nG\,Mpc^{1/2}$. The Auger exposure
corresponds to that calculated with purely geometrical considerations, which for
$E>4\times 10^{19}eV$ is probably not far from the truth. The result is
presented in Fig \ref{fig:2}. Galactic field has not been taken into account.
The most conspicuous object is the galaxy NGC 3256. It is a merging system -
"super starburst" galaxy [5] in a very vigorous episode of star formation, also
the most luminous X-ray starburst galaxy currently known. If the Arp 299 in the
northern hemisphere is responsible for the three events then Auger could
register about 4 showers from NGC 3256 for each 100 events above
$4\times10^{19}eV$. The expected number of doublets and triplets to be
registered by Auger for the LIRG hypothesis and isotropic sky is shown in Table
1 for 100, 200 and 300 total events. For higher number of showers it is looking
for individual sources rather than counting multiplets what will be a better way
of analysis. \begin{table}[t]
\begin{minipage}[b]{0.48\textwidth}
 \caption{Mean number of doublets and triplets (within separation angle
$\leq2.5^{\circ}$) predicted for Auger for LIRG and isotropic sky
($E>4\times10^{19}eV$).} \end{minipage} \hfill
\begin{tabular}{c|cc}
\hline
 & $<N_{doub}>$ &  $<N_{trip}>$  \\
$N_{events}$ & {\footnotesize LIRG/Iso} & {\footnotesize LIRG/Iso} \\
\hline
100  & $4.7/3.5$& $1.5/0.18$    \\
200     & $14/12.5$ &$5.6/1.3$ \\
300     &  $25/25$ & $21/8.3$ \\
\hline
\end{tabular}
\end{table}
\section{Conclusions}
The Auger experiment will soon see the southern sky with unprecedented
sensitivity (see eg.[2]), so that the hypothesis of the
luminous infrared galaxies as to be responsible for UHECR
particles (and  for the AGASA triplet) will probably be verified. The accuracy
of the shower direction determination in this experiment will be below one
degree. So, if any point-like excess is visible, its  correlation with an
object could be analysed by  finding the best Galactic and extragalactic
magnetic fields, as presented here.\par
\begin{footnotesize}
This work has been supported by Polish Ministry of Sci. Research and
Information Technology (grant no 2P03 011 24) and the University of Lodz grant
for Division Exp.Physics\\ \par
\end{footnotesize}
\re
1.\ Al-Dargazelli S.S. et al, \ 1996, J.Phys. G 22, 1825
\re
2.\ Bluemer J. for Auger Collaboration, \ 2003, 28th ICRC, Tsukuba
\re
3.\ Cronin J., \ 1997, Pierre Auger Design Report, http://www.auger.org
\re
4.\ Farrar G.R. and Biermann P.L., \ 1999, Phys. Rev. Lett. 83, 2472
\re
5.\ Joseph R.D. and Wright G.S., \ 1985, MNRAS, 214, 87
\re
6.\ Smialkowski A., Giller M. and Michalak W., \ 2002, J. Phys. G, 28,
1359
\re
7.\ Takeda M. et al,  \ 1999, ApJ, 522, 225
\re
8.\ Tinyakov P.G. and Tkachev I.I., \ 2002, Astropart. Phys., 18, 165
\re
\endofpaper \end{document}